\DocumentMetadata{}

\documentclass[nonacm, acmtog, screen]{acmart}

\usepackage{pgfplots}
\usepackage{tikz}
\usepackage{pgfplots}
\usepackage{pgfplotstable}
\pgfplotsset{compat=1.7}
\usepackage{subcaption}
\usepackage{xspace}
\usepackage{balance}

\newcommand{\LiveDB}{\textsc{LiveDB}\xspace}
\newcommand{\ArchiveDB}{\textsc{ArchiveDB}\xspace}
\newcommand{\StateDB}{\textsc{StateDB}\xspace}
\newcommand{\StateDBs}{\textsc{StateDBs}\xspace}

\AtBeginDocument{%
  \providecommand\BibTeX{{%
    \normalfont B\kern-0.5em{\scshape i\kern-0.25em b}\kern-0.8em\TeX}}}

\setcopyright{acmcopyright}
\copyrightyear{2018}
\acmYear{2018}
\acmDOI{XXXXXXX.XXXXXXX}

\acmConference[Conference acronym 'XX]{Make sure to enter the correct
  conference title from your rights confirmation emai}{June 03--05,
  2018}{Woodstock, NY}
\acmBooktitle{Woodstock '18: ACM Symposium on Neural Gaze Detection,
 June 03--05, 2018, Woodstock, NY} 
\acmPrice{15.00}
\acmISBN{978-1-4503-XXXX-X/18/06}

\usepackage{empheq}
\usepackage{xcolor}

\usepackage{tikz}

\newtheorem{example}{Example}

\usepackage{subcaption}

\begin{document}

\title{Efficient Forkless Blockchain Databases}

\author{Herbert Jordan}
\email{herbert@fantom.foundation}
\affiliation{%
  \institution{Sonic Labs}
  \country{Austria}
}

\author{Kamil Jezek}
\email{kamil@fantom.foundation}
\affiliation{%
  \institution{Sonic Labs}
  \country{Czechia}
}

\author{Pavle Suboti\'{c}}
\email{psubotic@fantom.foundation}
\affiliation{%
  \institution{Sonic Labs}
  \country{Serbia}
}

\author{Bernhard Scholz}
\email{bernhard@fantom.foundation}
\affiliation{%
  \institution{Sonic Labs}
  \country{Australia}
}

\begin{abstract}
Operating nodes in an L1 blockchain remains costly despite recent advances in blockchain technology. One of the most resource-intensive components of a node is the blockchain database, also known as \StateDB, that manages balances, nonce, code, and the persistent storage of accounts/smart contracts. 
Although the blockchain industry has transitioned from forking to forkless chains due to improved consensus protocols, forkless blockchains still rely on legacy forking databases that are suboptimal for their purposes. 

In this paper, we propose a forkless blockchain database that avoids costly operations associated with forking. Our improved database structure exhibits a 100x improvement in storage and a 10x improvement in throughput compared to the geth-based Fantom Blockchain client.
\end{abstract}



\maketitle

\section{Introduction}
\label{sec:introduction}
Blockchain technology is the core building block for democratizing finance.
At its essence, a blockchain can be perceived as a trust machine composed of a network of nodes that secure a shared ledger.
For the blockchain to operate, each node runs software that communicates with the other nodes in the blockchain.
This software consists of several essential components that accept transactions (transaction pool), agree on the next block (consensus protocol), execute transaction instructions (virtual machine), and a blockchain database that stores the blockchain worldstate (also known as \StateDB).

Early proof-of-work blockchains, such as Bitcoin and Ethereum, are hampered by limited performance, i.e., achieving less than 30 transactions per second (TPS)~\cite{ethtps}.
These blockchains have forking semantics; when the ledger's consistency cannot be temporarily enforced, multiple forks are maintained and processed using different versions of the account's state.
Eventually, a single fork is globally selected based on a heuristic, such as the longest-chain rule~\cite{Nakamoto2009BitcoinA}. 

In an effort to improve performance, modern blockchains have adopted forkless proof-of-stake consensus protocols.
In these block\-chains, the consensus protocol ensures that a consistent set of blocks is produced among the nodes, thereby avoiding forking.
The vast majority of modern blockchains incorporate proof-of-stake consensus protocols~\cite{10.1155/2022/2812526,tendermint,bullshark,pbft,mys,BairdL20,solana} to achieve a throughput of thousands of transactions per second~\cite{abs-1910-11143}. 

However, forkless blockchains continue to use legacy databases that were initially designed for forking blockchains only. 
Recent profiling experiences~\cite{abs-1910-11143} report that transaction execution in forkless blockchains is now primarily dominated by storage access, which consumes up to 75\% of the block processing.
Moreover, they require nodes to store terabytes of data\footnote{Storing the entire history of the Ethereum mainnet currently requires 24 TB of disk space, according to \url{https://etherscan.io/chartsync/chainarchive}}.

To better understand the problem, consider the popular Merkel Patricia Trie (MPT) data structure used as a foundational data structure in \StateDBs.
The MPT is an \emph{appending} data structure where each leaf node stores a value (e.g., a balance of an account, etc.), and the path from the root to the leaf node corresponds to a key (e.g., account address).
Each inner node in the MPT stores the cryptographic hash of the concatenated contents of all its children, which multiple worldstates of the \StateDB can use if the contents of the children have remained unchanged across different forks/blocks on the chain.
An MPT’s root hash serves as the commitment of the blockchain state for authentication and refers to a version of the worldstate.
Each worldstate has its root hash node.
Thus, the MPT is highly suitable for storing multiple versions of the worldstate that slightly differ.  

However, the MPTs ability to track multiple versions of a block state with multiple roots is costly as it requires pruning (i.e., disposing of worldstates that are no longer needed) and copying of data. 
And yet, for forkless blockchains, this behavior is superfluous.
As a result of this one-size-fits-all solution, forkless blockchains experience limited throughput and require nodes with terabytes of expensive storage. 
We demonstrate this phenomenon below in Example~\ref{ex:mptbad}. 

\begin{example}
\label{ex:mptbad}
Let us represent an MPT as a family of extensional functions $f_x: K \rightarrow V$ that are parameterized by version $x\in X$ from a version space $X$ with key set $K$ and value set $V$.
The keys address the accounts with their payloads.
The value set may represent concrete values for account balances, nonces, smart contract codes, or storage values. Consider the MPT on the left in Figure~\ref{fig:pruning}. This represents the MPT in an initial state. 
It consists of a single root $x_1$ that describes the following mapping: $k_1 \mapsto v_1, \ldots, k_4 \mapsto v_4$ where $k_1 = a \cdot c$, $k_2 = a \cdot d$, $k_3= b \cdot e$, and $k_4 = b \cdot f$.
If we execute a transaction, the mapping changes from $k_4 \mapsto v'_4$ as shown on the right-hand-side of Figure~\ref{fig:pruning}. Here we introduce a new root $x_2$ representing the new version and duplicate the path along edges $b$ and $f$. 

Note that the highlighted path for a forkless blockchain is redundant, as it is not needed for the new root $x_2$. Retaining the data causes bloat and thus will eventually be \emph{pruned}~\cite{livepruning}, which typically requires putting the node offline.  Moreover, as shown by the dotted arrows, numerous copies of keys along the path need to be copied to create a path from $x_2$ to the new value $v'_4$. Overall, considering an MPT typically holds hundreds of millions of values, the highlighted inefficiencies lead to considerable computational overhead and redundant bloat if previous versions are not needed. 

\begin{figure*}[htbp]
        \centering
        \includegraphics[width=0.45\textwidth]{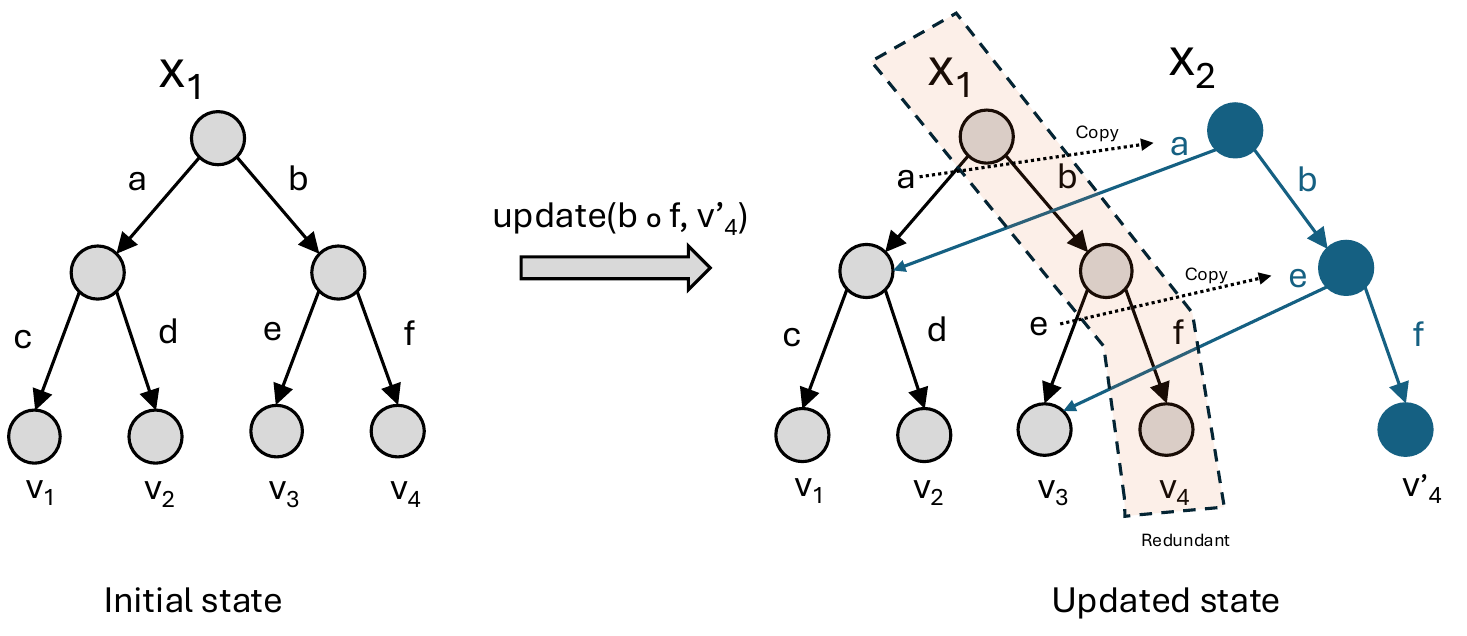}
        \vspace{-3mm}
        \caption{Evolution of Merkle Patricia Trie: A new version of state $x_1$ is constructed, resulting in a new state $x_2$ by updating the value $v_4$ to $v_4'$ for the key $b \cdot f$. \label{fig:pruning}}
\end{figure*}

\end{example}

In this paper, we argue that \emph{forkless blockchains need forkless blockchain databases}. We present a new state database design for forkless blockchains that provides high throughput and low storage costs compared to legacy alternatives. 

In our approach, we first specialize the state database according to its intended \emph{role} in the blockchain. For instance, the \LiveDB is a database for validator and observer roles, which require only the state for the latest block on the blockchain. The \ArchiveDB is a database for archival roles that keep all historical states for previous blocks. 

The key aspect of our database design is that the \LiveDB uses a mutable data structure that allows data to be \emph{overwritten}, reflecting the constrained versioning occurring in forkless blockchains. 
Overwriting allows us to reflect the state of the most recent execution without keeping stale states in the database (right MPT in Figure~\ref{fig:pruning}). 
Thus, we perform pruning at no additional cost with \LiveDB.
We refer to this method as \emph{intrinsic pruning} that is inherently performed as a consequence of the chosen data structure. 
Moreover, the \LiveDB stores key-value maps as densely as possible. A dense representation of the key-value maps results in smaller disk footprints and faster accesses. 
This design is in stark contrast to the MPT implementations that use key-value stores (e.g., LevelDB~\cite{leveldb}, RocksDB~\cite{rocksdb}) and SQL databases.

In the case of \ArchiveDB, we use an appending and monotonically growing data representation that can compress a forkless chain of blocks efficiently and thus reduce repeated copying of the keys (dotted arrows in right MPT in Figure~\ref{fig:pruning}) and re-calculation of hashes. This is a side effect of MPTs when value changes occur in the leaf nodes. The version dependencies in an MPT data structure form a tree (though suitable for the longest-chain rule in forked chains). The \ArchiveDB permits only a single subsequent version for a single block and, hence, can store the changes between two subsequent blocks more efficiently in a delta update style.

We have implemented \LiveDB and \ArchiveDB  in the Fantom Blockchain and evaluated their utility on real world transactions in a public testnet. Our evaluation shows that \LiveDB reduces the storage footprint\footnote{Note that even smaller storage decreases imply significant storage reductions linear in the number of nodes in the network because each node replicates the data in a classical L1 blockchain.} on the Fantom blockchain by approximately $100\times$ while increasing throughput $10\times$ compared to using the MPT-based Ethereum geth client~\cite{geth}. 

We summarize our contributions as follows:
\begin{enumerate}
    \item A novel design of \StateDB for forkless blockchains that separates \StateDB into \LiveDB and \ArchiveDB depending on the role (Validator vs Archive node),
    \item \LiveDB intrinsically provides live pruning with a new data structure, and
    \item \ArchiveDB is a linearly compressed database schema capturing the changes between two subsequent blocks without versioning.
\end{enumerate}

This paper is structured as follows: In Section~\ref{sec:dbspec}, we present our role-based database architecture. 
In Section~\ref{sec:live-db} and ~\ref{sec:arch-db}, we present our \LiveDB and \ArchiveDB specializations, respectively. In Section~\ref{sec:implementation}, we present the implementation details of both databases. 
In Section~\ref{sec:evaluation}, we evaluate both databases, and we present related work in Section~\ref{sec:related}, concluding in Section~\ref{sec:conclusion}. 

\section{StateDB Specialization}
\label{sec:dbspec}

    \begin{figure*}[h]
    \centering
    \begin{subfigure}{0.30\textwidth}
    \includegraphics[width=0.9\textwidth]{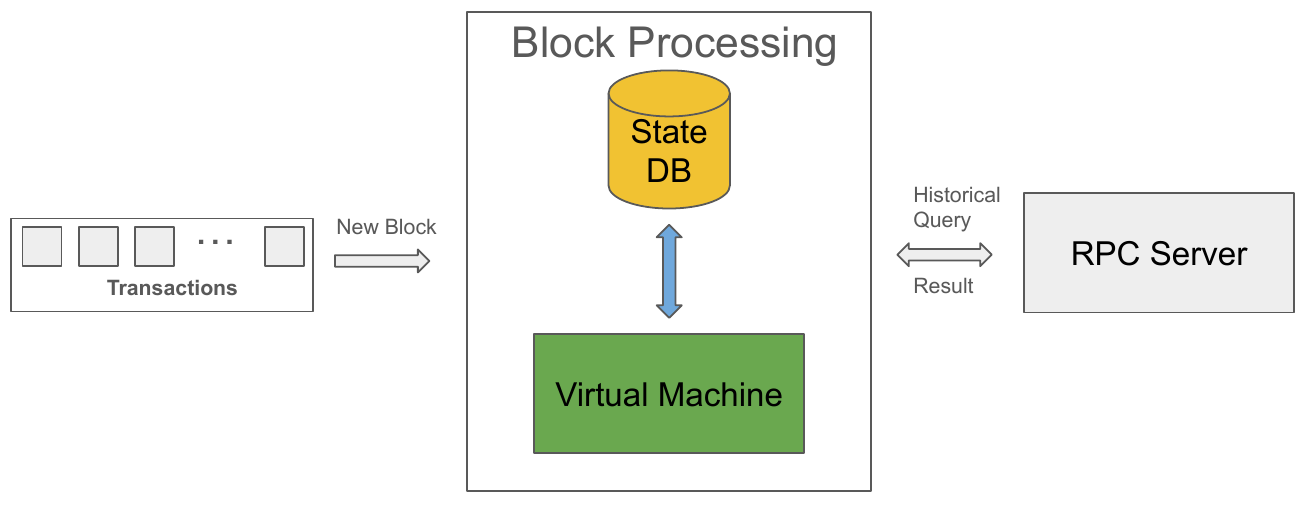}
    \caption{Unspecialized Database Design}
    \label{fig:trad}
    \end{subfigure}
    \hfill
    \begin{subfigure}{0.30\textwidth}
    \includegraphics[width=0.7\textwidth]{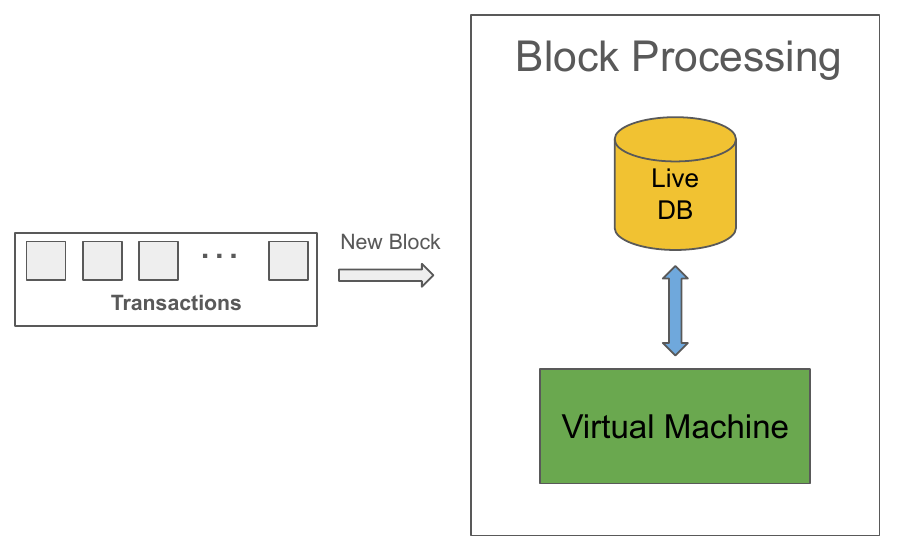}
    \caption{Specialized Database Design for Validators Nodes}
    \label{fig:live}
    \end{subfigure}
    \hfill
    \begin{subfigure}{0.33\textwidth}
    \includegraphics[width=0.9\textwidth]{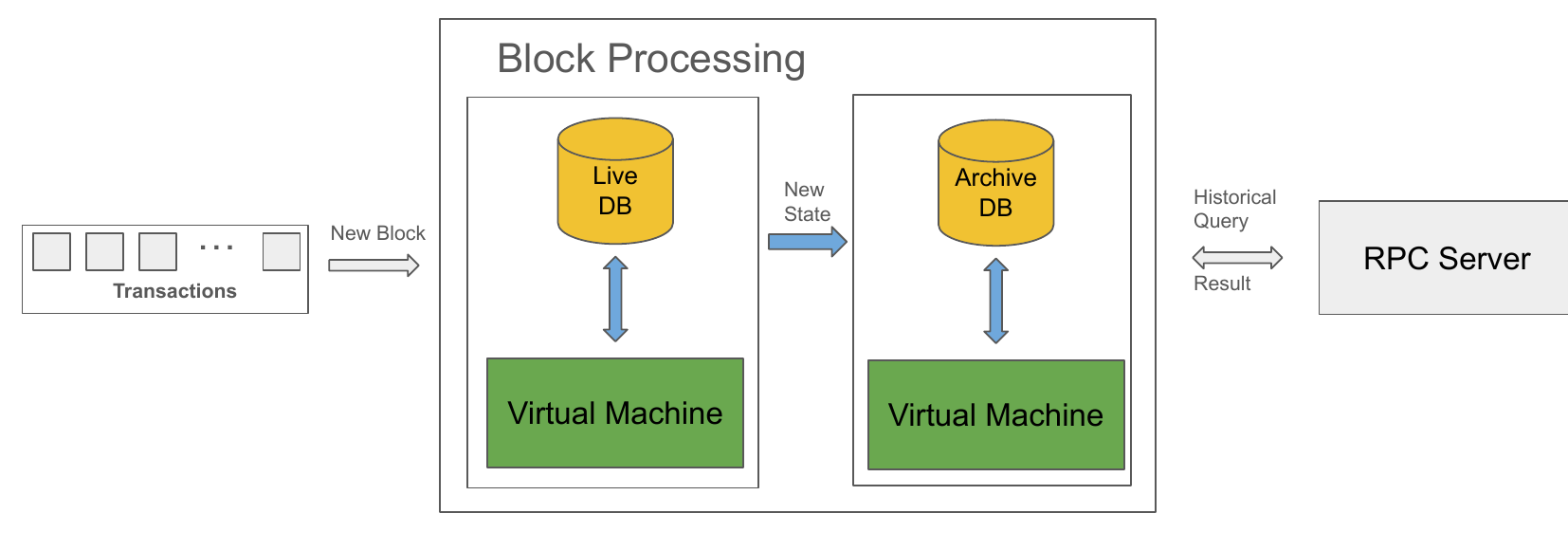}
    \caption{Specialized Database Design for Archive Nodes}
    \label{fig:archive}
    \end{subfigure}
    \caption{Blockchain Database Architecture: Unspecialized and specialized databases for the assigned roles of nodes.}
    \label{fig:tradvslive}
    \end{figure*}

Nodes in a blockchain have different roles.
A node is assigned a role by configuring the software, called the \emph{client}, that operates on each node.

Validator roles are tasked with producing a new block in the blockchain. 
To achieve this, validators run a proof-of-stake consensus protocol~\cite{POS}, which adds new transactions to the ledger and consistently orders them into blocks.
The block processor of the validator executes the transactions using a virtual machine and a blockchain database (\StateDB).

The \StateDB is necessary to verify the validity of transactions, including those that have been skipped due to insufficient funds, incompatible nonces, and other reasons.
Since the validator is tasked only with producing the next block, it does not need to retain any historical state; it only requires the latest state of the \StateDB. 
Therefore, after each processed block of transactions, the state of the \StateDB can evolve to the latest state.
The observer role is similar to the validator role, but it cannot insert new transactions into the ledger; instead, it participates in the consensus protocol and updates its state accordingly. The archive node role is an extended observer role that keeps the historical states of the blockchain, which can be queried via Remote Procedure Call (RPC). 
Examples of historical queries include querying an account balance at a specific block height or executing a smart contract at a given block height that queries the information only (but does not change the state). 
For this reason, it needs to keep the linear evolution of states in the \StateDB.

A typical block processing design is illustrated in Figure~\ref{fig:trad}, utilizing an unspecialized \StateDB design employed in chains such as Ethereum and Fantom~\cite{fantom}.
Here, the consensus component generates new blocks, enabling the block processor, which includes the virtual machine and the \StateDB, to advance to the next state.
In this block processing architecture, a single \StateDB is used to interact with both the Virtual Machine (VM) and external RPC calls for historical queries, such as querying the balance of an account in a specific block.
A standard implementation for a \StateDB is based on Merkle-Patricia Trie~\cite{merkle} that supports both archival and live data storage.
However, the configuration in Figure~\ref{fig:trad} results in excessive storage overhead for validator and observer nodes that do not need historical states.

To overcome the shortcomings of this one-size-fits-all \StateDB, we propose a new architecture for \StateDB, featuring two specialized databases: \LiveDB and \ArchiveDB. The two use cases of the databases are summarised in Figure~\ref{fig:live} and ~\ref{fig:archive}. 

In Figure~\ref{fig:live}, validators and observers only have a \LiveDB but no \ArchiveDB. The VM will access and update the \LiveDB when processing the next block. 
The \LiveDB maintains a single state for its accounts and has no multiple worldstates.

In Figure~\ref{fig:archive}, the \ArchiveDB is used to process historical RPC requests and is read-only by the RPC server. When a new block arrives, block processing generates new states using the \LiveDB and writes the updated accounts for a block into the \ArchiveDB.

\section{LiveDB  with Intrinsic-Pruning}
\label{sec:live-db}


\begin{figure*}[htbp]
    \centering
    \includegraphics[width=0.7\textwidth]{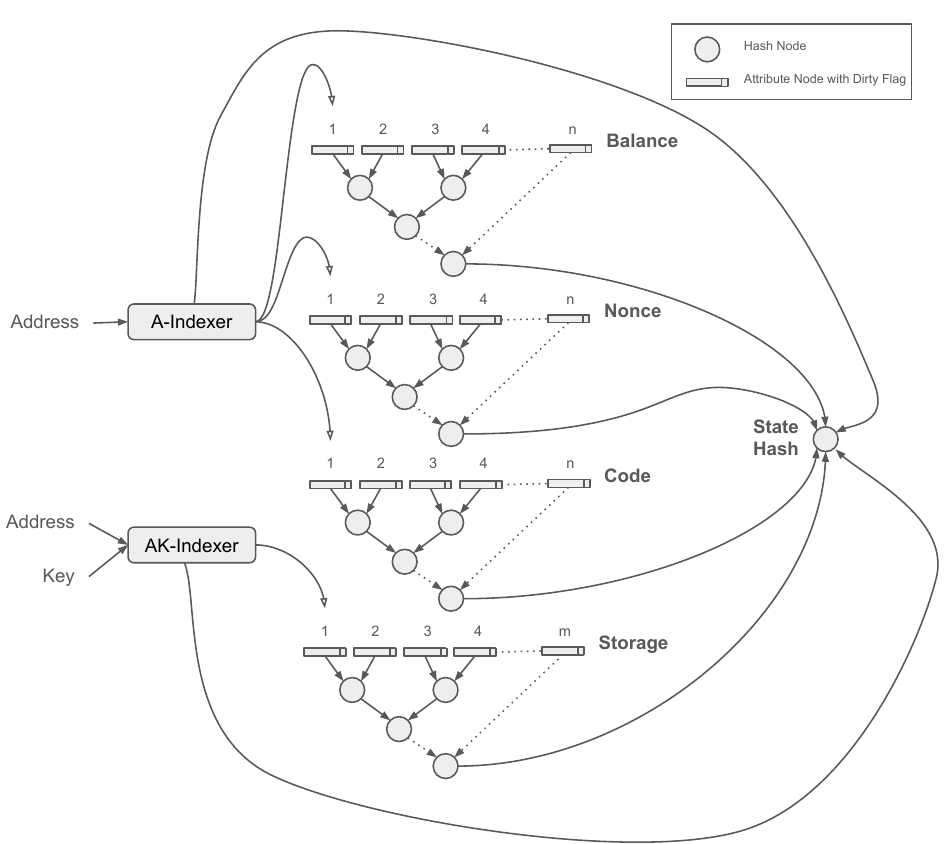}
    \caption{Hash calculation mechanism: Hashes are aggregated for attributes/storage values across all accounts, and indexers produce hashes for their stored keys.}
    \label{fig:concept}
\end{figure*}

In this section, we describe the design of the \LiveDB, which is designed for the validator and observer roles.

The \LiveDB design has two design principles. 
The first is a normalization~\cite{10.5555/551350} step where the attributes of accounts are not directly fetched. Instead, an account address is mapped to a record number, and with the record number, the account attribute is retrieved.
For smart contract storage, we use an account address and storage key mapping that maps to the record number of the storage value.
This allows for the direct storage of account attributes and their corresponding values.
Instead of using a B-Tree or other ordered data structures with ${\mathcal O}(\log n)$, we use an efficient I/O hashmap~\cite{linear-hashing} for the normalization mapping account addresses/storage keys to record numbers.

The second design principle is to use binary files that store account and storage attributes in fixed-length records. With a record number, an account attribute, or a storage value can be fetched in constant time by seeking the record and reading the record from the binary file.
The seek position is calculated using a linear function over its record number as an argument and the size of the attribute. In \LiveDB there is a separate file for each attribute to permit the direct access of attributes via their record number. For the code attribute, we require an extension to fixed-length records, as the code of a smart contract varies and its maximum size is approximately 25 KB.

The normalization is visualized in Figure~\ref{fig:concept}, which shows the two hashmaps.
The first hashmap \textsc{A-Indexer} maps an account address to a record number. 
With the record number, the account attributes (including Balance, Nonce, Code) of an account can be accessed directly in constant time.
The attributes of an account are stored under the same record number since the record number is a property of the account address.
The second hashmap, \textsc{AK-Indexer}, maps an account address and a storage key to a record number for accessing a storage value. 


There are hashing strategies, such as linear hashing~\cite{linear-hashing}, that are I/O-friendly, allowing for efficient hash-table growth via splitting without requiring a complete reconstruction of the hash table. 
This is achieved by expanding the bits of the index calculation depending on the hash table size. 
The hash table stores the keys and their corresponding record numbers, allowing for the retrieval of account records or storage values.
If the stored key in the hashtable is not identical to the searched key, the subsequent entries are searched until the key is found. 
The hash table stores the account address/storage key and the record number of the attributes/storage value as shown in Figure~\ref{fig:indexer}(b). 
The key is needed so that, in the event of a collision caused by the hash function, we can find a neighboring slot for storing the key and its ordinal number.


\subsection{Hash Calculations}

Hash calculations for the worldstate in the \LiveDB are performed lazily.
If an account attribute or a storage key of an account changes, the attribute will be marked as dirty, as illustrated in Fig~\ref{fig:concept}, but the overall hash will not be computed eagerly.

When a request for a hash calculation of the state is issued, the hash calculation is performed partially for the changed attributes and changed storage values. 
The actual hash calculations are split into several components.
Each account attribute has a unique hash value across all accounts.
The storage values across all accounts also have a hash value, and the indexers containing the account addresses and keys have their own hash values.
The hashes of these components are combined using a hash function, resulting in a single 32-byte hash value that represents the state root hash of the database.

For the file-mapped arrays of attributes and smart contract storage values, an underlying balanced binary tree is constructed for hash calculations using the storage/attribute values in a binary file.  
We refer to the balanced binary trees as the \emph{hash trees} of the attributes and storage values.
They are also illustrated in Figure~\ref{fig:concept}.
We can construct balanced binary trees using ideas from the heap data structure, as employed in heap sorts and other algorithms~\cite{10.1145/512274.512284,10.5555/1614191}, by utilizing arrays and index calculations to represent parent-child relations.
Hence, hash trees can also be implemented as file-mapped arrays.
Additionally, the balanced hash tree can be incrementally extended to accommodate new attribute/storage value entries without compromising the integrity of large portions of the hashes. 
Whenever a new attribute is inserted, it is appended at the end of the array, consequently creating a new branch of the hashing tree siting above this array. The array itself is always expanded by chunks to accommodate branching factor of the hashing tree, and the new chunk is created when the previous one is filled. This new chunk builds a new level of the hashing tree appended at the top of the tree, thus only this part of the tree must be recomputed.

A leaf of the hash tree corresponds to an attribute.
Inner nodes of the hash tree represent the aggregation of hashes of the previous layer.
The root node of the hash tree represents either the hash of an account's attribute or the storage hash over all accounts.
In contrast to MPT, the size of the hash tree is ${\mathcal O}(n \log_2 n)$ and the hashes are densely stored. 

The advantage of a lazy update is that for a set of changed attributes/storage values, the hash computation is performed in one go by visiting a node in the hashtree at most once.
This improves the amortized complexity and is achieved by using a reverse breadth-first traversal, which can be fully parallelized using the ideas of prefix-sum~\cite{10.1145/322217.322232}. 

\begin{figure*}[htbp]
    \centering
    \includegraphics[width=0.6\textwidth]{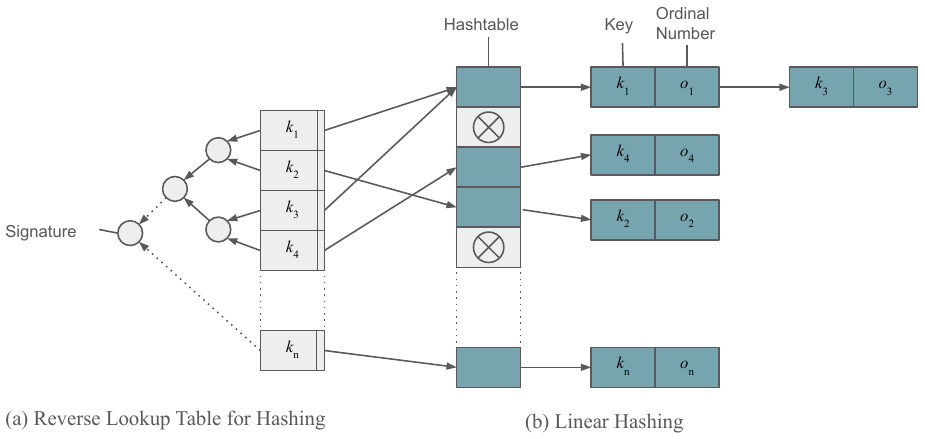}
    \caption{Indexer: An indexer consists of a hash table that maps account addresses/storage keys to record numbers and a reverse lookup table that stores the addresses/keys file-mapped array permitting dense hash calculations on the reverse lookup table.}
    \label{fig:indexer}
\end{figure*}

The hash calculations for the HashMap A-Indexer and AK-Indexer are more complex, as the hash table may contain holes that prevent a dense hash calculation, as previously outlined for account attributes and storage values. 
As shown in the Figure~\ref{fig:indexer}(a), we extend the indexers with a reverse lookup table.
It is also a file-mapped array that permits dense hash calculation for account attributes and storage values.
The index $i$ of entry $k_i$ corresponds to the record number of the account address/storage keys. 
The entry $k_i$ is either an account address for A-Indexer or an account address/storage key pair for the AK-Indexer.
The keys are stored linearly; hence, the same balanced binary hash tree can be constructed for the reverse lookup table. 
When a storage value/attribute with a new address/storage key is added, the indexer appends the new address/key at the end of the reverse lookup table. 
Note that if the address/storage key already exists, the reverse lookup table of the hash map remains unchanged.
We have only new entries for new address/storage keys in the reverse lookup tables.

For our \LiveDB we have an underlying assumption that must not be broken.
Appending new account addresses/storage keys must occur in the same order on all nodes, and we assume that the construction of the world state is identical on all nodes. 
If this assumption is broken, the hash of two network nodes may diverge since the order of keys may differ, and hence the hash of the indexers may differ between two different nodes. 
Note that this is not a strong assumption since all nodes must have deterministic computations to replicate the state in the network consistently.
However, the MPT data structure does not have this assumption.
The order of construction of the world state will not impact the root hash.
This has implications for the exchange of states between nodes, i.e., the keys of the indexer must be exchanged in order, and no arbitrary order can be assumed for exchanging the keys of an indexer.
Also, the key order of the reverse lookup tables becomes a weak certificate for the computations that constructed the worldstate.

\subsection{Optimizations}
\begin{figure*}[htbp]
    \centering
    \includegraphics[width=0.6\textwidth]{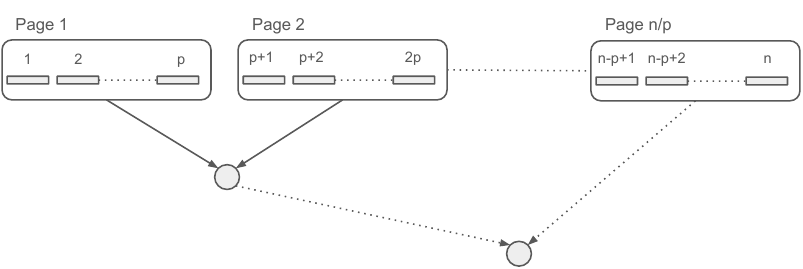}
    \caption{Paging mechanism: $\frac{n}{p}$ pages where $p$ elements of an attribute are packed into a page.}
    \label{fig:paging}
\end{figure*}

We improve the storage complexity by a constant factor $p$ by aggregating the attributes/storage values to pages of fixed size, resulting in $\frac{n}{p}$ pages with $p$ elements per page.
Consider Figure~\ref{fig:paging} that packs $p$ elements of an attribute into a page.
Instead of hashing an element, a whole page is hashed, reducing the size of the hash tree. 
Let's assume that $p=2^l$; the number of leaves in the hash tree is reduced to $\frac{n}{2 p}=2^{k-l-1}$ and the hash tree size becomes $\sum{0\leq i \leq k-l-1}2^i=2^{k-l}-1 = \frac{n}{p}-1$, i.e., the reduction of the hash tree is linear in the number of slots inside a page. 
The page size $p$ is a classical computational-space trade-off. 
As the page size increases, more computations are required to calculate the hash of an attribute, since the number of slots per page also increases.
However, the number of hash nodes in the hash tree reduces.

\section{ArchiveDB for Archival Nodes}
\label{sec:arch-db}
In this section, we describe the design of the \ArchiveDB, which is specialized for the archive role. 

The \ArchiveDB  has the same data model as \LiveDB, though it is indexed by a block number that permits the retrieval of a historical block state.
The central design concept of \ArchiveDB is to store historical state data in sorted tables, building a log of changes over the block numbers for each account and storage keys. 
Log entries are created if attributes/storage values change.
\ArchiveDB can employ logs because there are no forks, and for a single block, there can be at most one change of an attribute/storage value.
Efficient data structures, such as Log-Structured Merge Trees, are suddenly amenable to forkless chains, although they are not applicable in forked chains.

For instance, to retain historic data on the state of storage, defined by an account and a storage key, a table of tuples of the format
\begin{verse}
[\emph{account}, \emph{key}, \emph{block}, \emph{value}]
\end{verse}
is constructed by keeping track of the changes that occur. 
Each time a storage key is mutated in a block for an account, a new tuple with the current block number is added. 
Thus, a log of mutations is gradually recorded.  
By storing those tuples in a sorted order, a whole history track of mutations can be created. This is only possible because we have a forkless chain.
For each block, there exists at most one version of a value.

For example, the content of the sorted table may look like this:
\begin{center}
\begin{tabular}
{|l|l|l|l|}
\hline
Account &
Key & 
Block &
Value \\
\hline
$\ldots$ & 
$\ldots$ & 
$\ldots$ & 
$\ldots$ \\
\texttt{0x123} &
\texttt{0x00001} &
\texttt{14} &
\texttt{100} \\
\texttt{0x123} &
\texttt{0x00001} &
\texttt{16} &
\texttt{110} \\
\texttt{0x123} &
\texttt{0x00004} &
\texttt{14} &
\texttt{80} \\
\texttt{0x123} &
\texttt{0x00004} &
\texttt{17} &
\texttt{90} \\
\texttt{0x140} &
\texttt{0x00001} &
\texttt{8} &
\texttt{100} \\
$\ldots$ & 
$\ldots$ & 
$\ldots$ & 
$\ldots$ \\
\hline
\end{tabular}
\end{center}

To determine, for instance, the value of storage key \texttt{0x00001} of account \texttt{0x123} at block \texttt{15}, we issue a search query that finds the first entry whose account and storage keys match and whose entry is smaller than or equal to the block number \texttt{15}.
The search query in this example would return the entry \texttt{[0x123, 0x00001, 14, 100]} since this is the value of the account/key that would have been visible in block \texttt{15}.

For the search query, we require additional checks to determine whether the found entry in the log describes the storage location and account of the state query, and whether the account has been deleted before the block. 
However, the basic design of \ArchiveDB is to reduce the lookup of historic information to sorted table lookups.
Besides the table for storage values, tables for historical balances, nonces, codes, and account states (both existing and deleted) are required to complete the full historical record. 
However, these tables have one column less. 
For example, the table for balance would look like:
\begin{verse}
[\emph{account}, \emph{block}, \emph{balance}]
\end{verse}

The key advantage of using logs instead of MPTs is that MPTs require the storage/re-calculation of hashes from the leaf that stores the attribute/storage value to the root node. 
This results in additional storage overhead and work, compared to using logs only.

The implementation of \ArchiveDB is performance-oriented, and log entry writing is implemented as a background task.
When a new block is committed, the differences to the previous worldstate are computed and used for writing the logs.
These logs are then fed asynchronously into the \ArchiveDB's tables using a background thread. 
\ArchiveDB can serve queries and append new data simultaneously, mainly because the new log entries never overwrite the old data.
Hence, no locking is required for existing records.
Existing entries in \ArchiveDB's tables will never be altered and stay immutable.

\subsection{Hash Calculations}
The hash computation mechanism in \ArchiveDB is an incremental process that uses the previous hash of the previous block and the attributes/storage values that have changed in the current block. 
Since only a small fraction of accounts are changed in a block, the hashing of the differential state of a block is performed per account.

To ensure consistency and determinism in hash generation, all account modifications within a block are ordered lexicographically by account address. 
Furthermore, the sequence of changes—such as updates to balance, nonce, and storage slots—is strictly maintained in a predefined order.
In cases where an account's storage is modified, the associated storage keys are also sorted lexicographically for hashing.

More technically, the hash of an account and a new block is computed by combining all changes using the byte serialization into a string, which is then hashed to produce a change-specific hash.
If the account has a prior state, its previous hash is retrieved and combined with the current update hash to generate a new composite hash for the account. This procedure is applied iteratively across all accounts affected in the current block.
The resulting account-level hashes are then aggregated to compute a singular hash representing the entire block.

These hashes are stored independently for each account and block, facilitating efficient retrieval and recomputation.
When determining the current hash of an account or block, the system combines the existing hash with the hashes of recent updates, thereby streamlining the verification and lookup processes.

The hashing mechanism allows for checking consistency of the \ArchiveDB at any block height given  ${\mathcal O}(1)$ access time complexity, while proving particular values of accounts and storage slots need a linear scan of the dataset.

\section{Implementation}
\label{sec:implementation}
In this section, we outline the implementation details of our approach.

The \LiveDB and \ArchiveDB have been implemented using Golang.
For the \LiveDB, we explored three different variants: an in-memory version, a file-based version, and a LevelDB version.
Each component and development phase underwent rigorous benchmarking to ensure optimal performance.
We used the built-in Golang benchmarking capability or processed sample transactions from the real blockchain.
The in-memory database served as our performance upper bound, representing the ideal scenario we aimed to approximate in real-world applications closely.
Conversely, the LevelDB database functioned as the performance lower bound, providing a baseline from which we aimed to achieve significantly higher performance compared to current commodity software.

The implementation of the database necessitated the use of several fundamental data structures, including hash maps, arrays, and caches.
The process was relatively straightforward for both the in-memory and LevelDB variants.
Specifically, the in-memory database leverages standard Golang maps (associative arrays), simplifying the design by allowing direct mapping of account addresses to accounts and storage keys to storage slots.
A generic structure was created to wrap \textit{map[K]V}, enabling instantiation for all key/value pairs to be stored.
Notably, no caches were required for the in-memory variant due to its inherent nature.
Similarly, the LevelDB database is designed for storing key/value pairs, which facilitated direct utilization in our implementation.

Our custom file-based implementation required the development and composition of several fundamental primitives.
Unlike in-memory and LevelDB variants, where the implementation was relatively straightforward, our custom approach necessitated the creation of bespoke primitives:

\begin{itemize} 
    \item \textbf{PagePool}: A structure that allows access to binary pages based on their consecutive ID. 
    The PagePool maintains a configured number of pages in memory, persisting pages to disk when they are evicted from memory and loading pages from disk to memory upon access. 
    \item \textbf{Array}: An indexable persistent structure that contains values at specific indexes. 
    It utilizes the PagePool to read and load array segments in chunks from the disk. 
    \item \textbf{Linear Hash}: A persistent structure designed to map keys to values. It leverages the PagePool to persist its segments with an in-memory overlay. 
    \item \textbf{HashTree}: A structure that operates on the PagePool to compute Merkle proofs\cite{merkle} from the pages. 
    \item \textbf{LruCache}: A cache implementing a least recently used (LRU) eviction policy, constructed using a map and a linked list.
\end{itemize}

The integration of these primitives has enabled the implementation of the database. 
The \textit{A-Indexer} and \textit{AK-Indexer} employ \textit{Linear Hash} to generate an index that converts account and storage keys into ordinal numbers.
These numbers are subsequently used as indices in the \textit{Array} to retrieve the final value.
The \textit{HashTree} provides the cryptographic hash over the dataset. 
The \textit{PagePool} ensures that frequently accessed data remains in memory.
However, when data is dispersed on the disk, frequent page evictions can occur.
To mitigate this, we implemented the \textit{LruCache} in front of each indexer to cache the most frequently used keys, enhancing overall performance. 

The implementation of the Archive was straightforward, utilizing commodity software. For the SQL variant, we employed SQLite\footnote{https://www.sqlite.org}, and for the Key/Value variant, we used a Golang wrapper for LevelDB\footnote{https://github.com/syndtr/goleveldb}.
These variants were selected for their minimal operational costs.
In both cases, the libraries were linked to the Go project, eliminating the need for additional installations.

For SQLite, we established the database schema during program startup using \textit{CREATE SQL} statements and encoded all necessary SQL statements for reading and appending data within the code using prepared statements to ensure maximum performance.
Since LevelDB is a schema-less database, no extra configuration was required for its immediate use.

Due to the modular architecture, different database variants can be easily swapped.
Notably, \LiveDB can be combined with the original MPT database implementations as an Archive to ensure full compatibility.
This enables data provision via RPC to clients that require features such as block hashes, storage proofs, and other functionalities compatible with original Ethereum-compatible systems.

\section{Evaluation}
\label{sec:evaluation}
We conducted a series of experiments to evaluate transaction throughput and the disk storage space required by various implementations of our technique.
For this purpose, we adapted the tool for blockchain recording and replaying \cite{Kim2021off-the-chain} to record blockchain data available on the Fantom mainnet.
This dataset comprised 50 million blocks, which were stored locally on an SSD NVMe disk.
The tool was subsequently used to reprocess these blocks through selected components of the Fantom client, thereby reconstructing the blockchain.
Specifically, we modified the block processing module, which encompasses the Ethereum Virtual Machine (EVM) and the database, to isolate the influence of other components, such as the network, consensus protocol, and transaction pool. 
This allowed us to precisely measure the number of transactions processed within a sampled time frame and the corresponding disk space consumed.

For all experiments, we utilized the same virtual machine to process blockchain transactions, systematically switching the database to use the original Merkle Patricia Trie (MPT), our newly developed file-based LiveDB, and both SQL and Key/Value-based archives. 
The data were stored locally on the NVMe disk, and we assessed both the transaction throughput and the disk space utilized.

\begin{description}
	\item[\textbf{\em Claim-I: Throughput Improvement.}] \hfill 
    
    We investigate the throughput of the various variants of our new system and compare them to the state-of-the-art Geth Ethereum client previously used in the Fantom blockchain.

	\item[\textbf{\em Claim-II: Storage Overhead Improvement.}]\hfill
    
    We investigate the disk space requirements of the various variants of our new system and compare them to those of the state-of-the-art Geth Ethereum client previously used in the Fantom blockchain.  
\end{description}

\subsection{Experimental Setup}
\paragraph{Hardware} 

Experiments were conducted in cloud environments utilizing Google Cloud as the infrastructure provider. 
Virtualized servers were provisioned with an n2-standard-16 configuration, featuring 16 virtual CPUs and 64 GB of memory.
The machines were equipped with local NVMe storage, offering the highest available performance.
The system ran on Debian Linux with Go version 1.23.

\paragraph{Experimental Data}

We employed the tool \cite{Kim2021off-the-chain}, which is designed to operate in two distinct phases.
In the first phase, the tool synchronized an existing blockchain and recorded the necessary input data to enable the re-execution of all transactions in each block.
This process was applied to the Fantom mainnet, a blockchain that has been operational since 2018. Notably, this recording phase was not included in the performance measurements.
In the second phase, the tool leveraged the recorded data to replay the entire blockchain at an accelerated rate, wherein the software and hardware imposed the only constraints.
This approach provides insights into key performance metrics, including the number of transactions processed per second and the volume of blockchain gas consumed over a given period.

\paragraph{Baseline} 

Two sets of experiments were conducted, both utilizing the client, which operates on the Fantom mainnet.
The initial implementation of the client incorporated a database layer derived from Ethereum’s Geth client, which internally relies on the LevelDB key-value database.
In the first experimental run, the system was executed using this original configuration.
Subsequently, the experiments were re-executed using the client with an alternative database layer, implemented according to our new design.
This approach enabled a comparative analysis of the performance characteristics of the two database architectures.

\subsection{Throughput Improvement Evaluation}

To evaluate the impact of our new technique on transaction processing speed, we conducted an experiment in which we replayed the previously recorded blockchain data while systematically swapping the underlying database implementations.
During this process, we periodically sampled the block height and computed the transactions per second (tx/s) at each interval.
The results were subsequently plotted in a chart to compare the performance of the different database approaches.

\begin{figure*}
    \centering
    \includegraphics[width=0.7\textwidth]{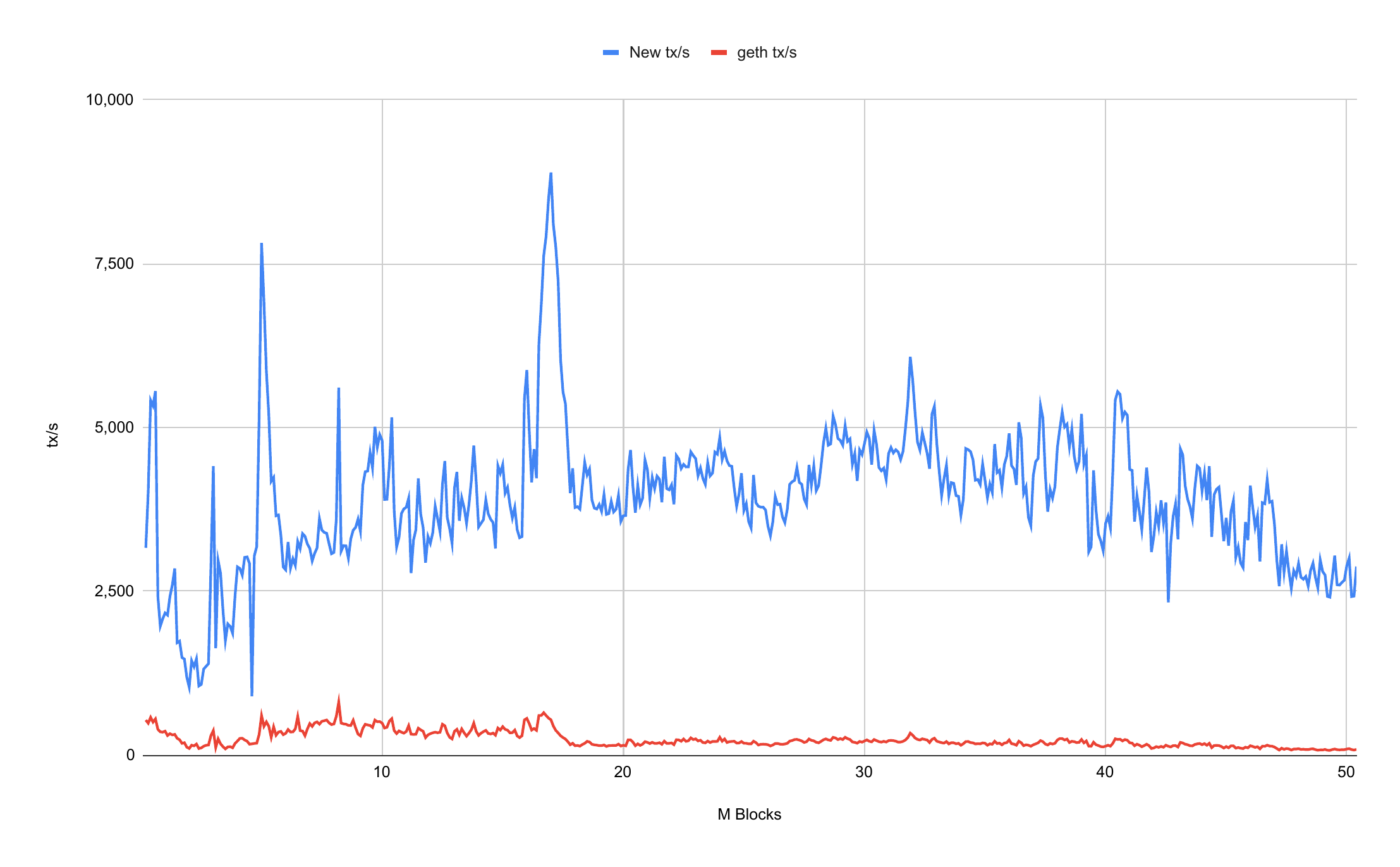}
    \caption{Transaction throughput comparison between original database (geth) and our New system. The new system is more than 10 times faster, featuring ~5.000 tx/s vs. below 500 tx/s for the MPT. }
    \label{fig:file-eval}
\end{figure*}

Figure~\ref{fig:file-eval} presents a comparative analysis of transaction throughput between the original Merkle Patricia Trie (MPT) and our novel scheme, which stores data directly in binary files.
The original MPT demonstrated a transaction processing speed of approximately 500 transactions per second (tx/s).
In contrast, our new method consistently achieved a transaction throughput of approximately 5,000 transactions per second, indicating a tenfold increase in performance.

Both databases exhibited a slight decline in performance after processing 40 million blocks.
This decline can be attributed to the increasing size of the database, which causes cache efficiency to diminish and necessitates more input/output (I/O) operations.
Despite this, the overall transaction processing speed remained relatively stable throughout the experiment. 

\begin{figure*}
    \centering
    \includegraphics[width=0.7\textwidth]{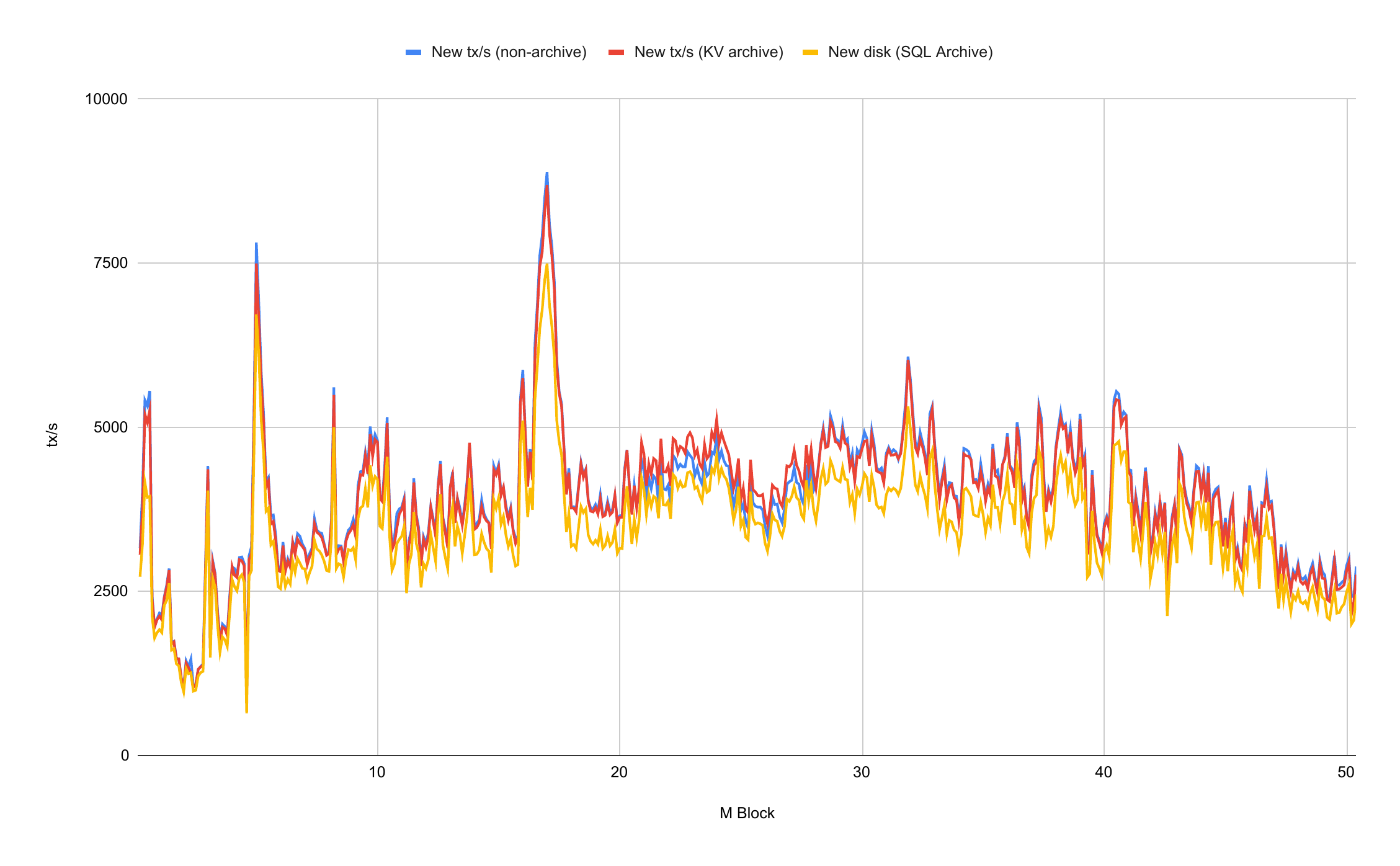}
    \caption{Throughput comparison of \LiveDB and two variants of \ArchiveDB; the Archive does not incur any significant slowdown.}
    \label{fig:carmen-archive}
\end{figure*}

Our design incorporates the capability to enable and swap the \ArchiveDB, which introduces the potential drawback of requiring data to be propagated twice to store it in both \LiveDB and the \ArchiveDB. 
To assess the impact of this on transaction throughput, we conducted a repeat of our previous experiment, this time enabling two variants of the archive: SQL and Key/Value databases.
We monitored transaction throughput once again and verified that no significant slowdown occurred. 
The results are presented in Figure \ref{fig:carmen-archive}, which plots \LiveDB alongside both archive variants.
As evident from the figure, the performance lines for \LiveDB and the two archives almost overlap, indicating minimal performance degradation.
The progression observed is comparable to that shown in Figure \ref{fig:file-eval}, where only \LiveDB was enabled.
From this experiment, we can conclude that enabling the archive does not adversely affect performance.

\subsection{Storage Overhead Improvement Evaluation}

In our investigation of various database schemes, we conducted an experiment involving the insertion of 50 million records into the database.
Upon completion, we utilized standard operating system tools to assess the size of the final directory where the database was stored. The results are presented in Table \ref{tab:database_sizes}.
When comparing the pruned MPT to our \LiveDB, the storage requirements were 1600 GB versus 30 GB, respectively.
This demonstrates that our new design conserves approximately 98\% of disk space.
In the case of the archive database, which required a substantial 17 TB when backed by MPT, our design required only 550 GB and 160 GB for SQL and Key/Value databases, respectively—again reflecting a reduction of around 98\% in storage space.

It is particularly noteworthy to highlight the size of the MPT archive.
The substantial 17 TB requirement may pose a limiting factor for certain operators entering the blockchain space, especially given that the use of relatively expensive SSD NVMe hard drives is recommended for the efficient operation of modern blockchains.
This significant storage demand underscores the challenges associated with scalability and accessibility for smaller enterprises or individual operators.
Our new archive design effectively addresses this issue by significantly reducing the required storage space, thereby enhancing the scalability and accessibility of the blockchain system for a broader range of users.

\begin{table}[h!]
    \centering
    \begin{tabular}{|l|c|l|c|}
        \hline
        \textbf{LiveDB} & \textbf{Size (GB)} & \textbf{Archive} & \textbf{Size (GB)} \\
        \hline
        MPT Pruned & 1600 & MPT & 17000 \\
        \hline
        LiveDB & 30 & SQL & 550 \\
        \hline
        & & Key/Value & 140 \\
        \hline
    \end{tabular}
    \caption{Comparison of LiveDB and ArchiveDB Sizes @ 50M blocks}
    \label{tab:database_sizes}
\end{table}

\subsection{Determining Optimal Pagesize}

To determine the optimal page size for \LiveDB, we conducted a series of practical experiments. 
The results indicated that a page size of four kilobytes offers superior performance, which aligns with the fact that contemporary hardware and operating systems are typically optimized for reading and writing data in blocks of this size.

\begin{figure*}[htbp]
    \centering
    \includegraphics[width=0.6\textwidth]{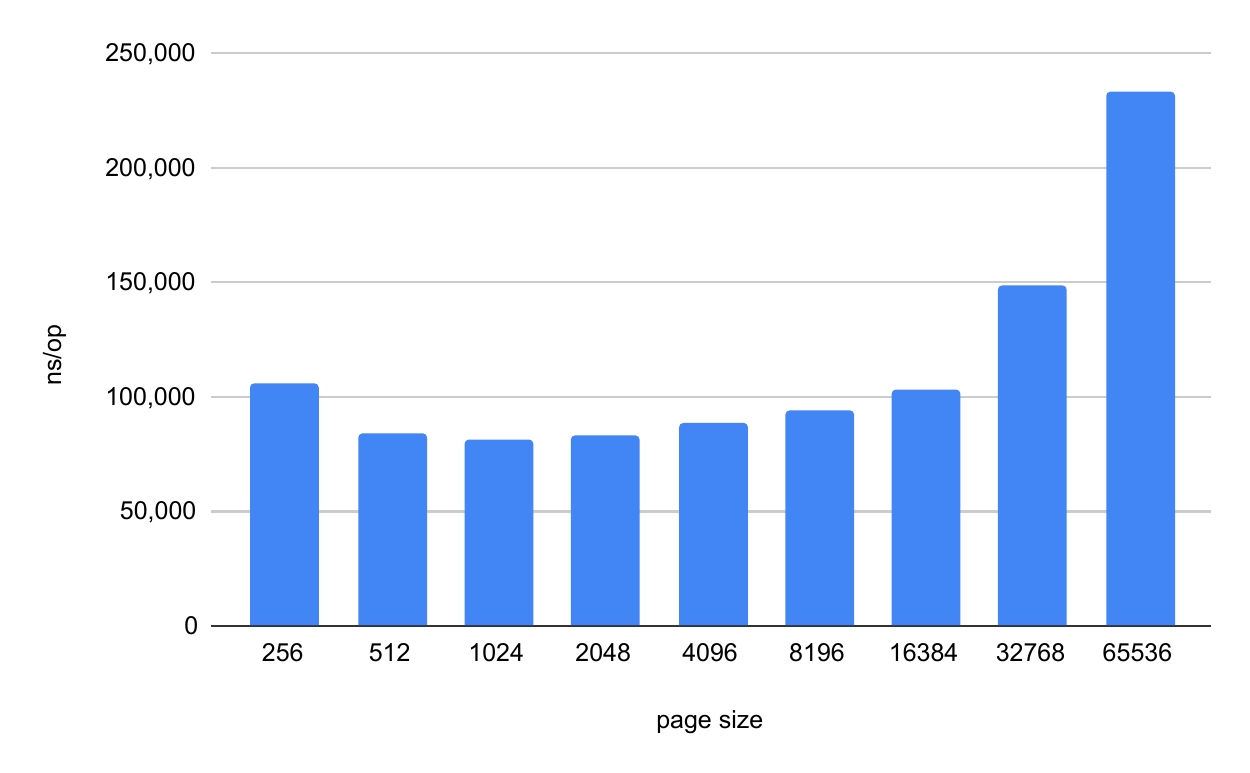}
    \caption{Time in nanoseconds to compute hash for varying page sizes on \LiveDB in memory using microbenchmark; smaller page sizes are better.}
    \label{fig:page-size}
\end{figure*}

The first experiment was a micro-benchmark designed to evaluate in-memory performance.
We inserted 16 million keys into the in-memory pages and repeatedly updated a fixed set of 100 items.
After each update, we computed a hash of the entire hash tree and measured the time required for this operation.
To ensure statistical reliability and minimize noise, we leveraged Go’s built-in microbenchmarking framework, which automatically repeats the experiment until performance stabilizes.
This benchmark was executed across a range of page sizes, and the results are presented in Fig. \ref{fig:page-size}, where the x-axis denotes page size and the y-axis shows the time in nanoseconds required to compute the hash.
The data reveals that smaller page sizes consistently yield better performance, supporting the hypothesis that large pages are inefficient in this context. 
Specifically, updating a single key within a large page invalidates a substantial volume of data, thereby increasing the cost of re-hashing.

\begin{figure*}[htbp]
    \centering
    \includegraphics[width=0.6\textwidth]{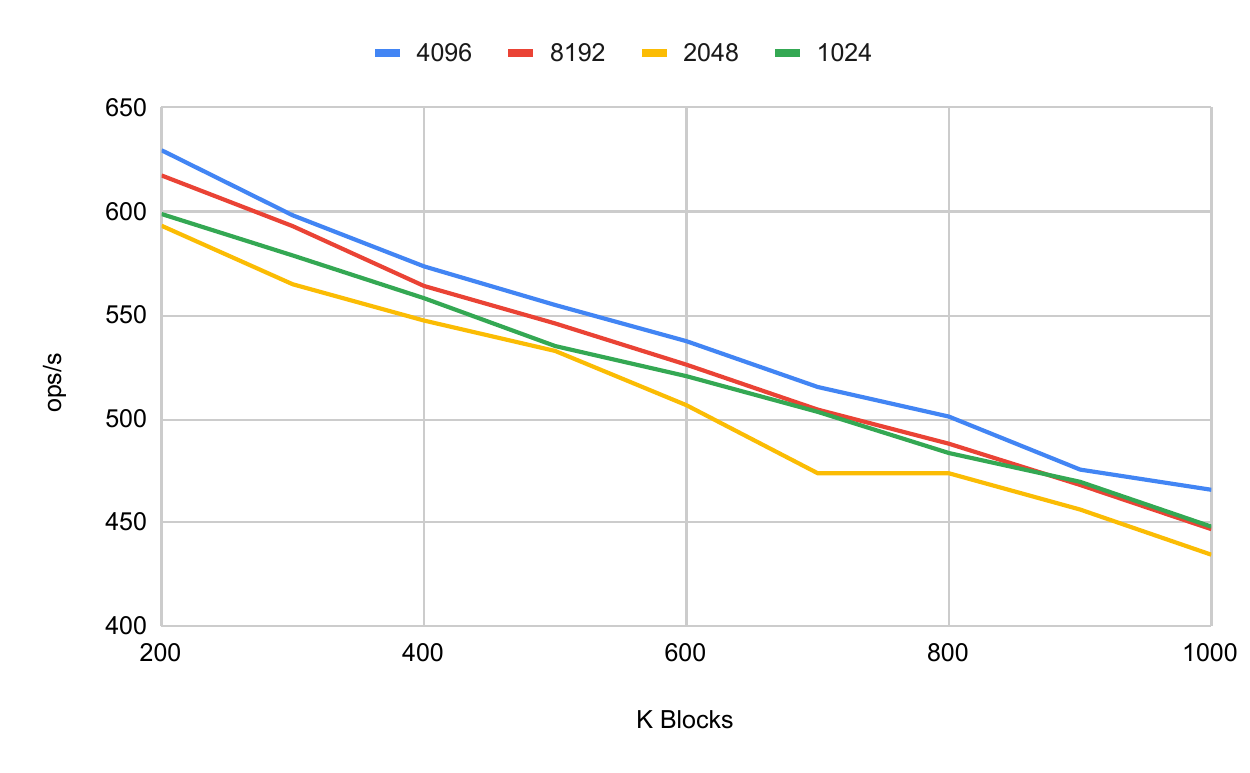}
    \caption{Number of hashes computed per second for varying page sizes utilizing I/O. Legend shows page sizes. The page size 4096 is optimal.}
    \label{fig:page-size-io}
\end{figure*}

To account for the potential impact of disk I/O, we conducted a second experiment using a virtual machine constrained to 8 GB of RAM, thereby forcing the system to utilize disk storage. We extended the dataset and repeated the benchmark for page sizes of 1024, 2048, 4096, and 8192 items—values identified as promising in the micro-benchmark.
The results, shown in Fig. \ref{fig:page-size-io}, indicate that while performance differences were less pronounced under I/O constraints, the page size of 4096 items consistently delivered the best results.

\section{Related Work}
\label{sec:related}

The Merkle Patricia Trie (MPT) is a fundamental data structure in blockchain systems, combining the advantages of Merkle trees and Patricia tries.
Merkle trees, introduced by Merkle~\cite{merkle}, provide a cryptographic method to ensure data integrity by hashing values of nodes in the trie.
Patricia, as proposed by Morrison~\cite{Morrison68}, offers an efficient method for compressing sparse data by compressing common paths of keys into shared sequences. 
Read and write amplification in tries has been a significant research focus. Raju~\cite{raju2018mlsm} introduced the Merkelized LSM, integrating Merkle and log-structured merge tree data structures. Ponnapalli~\cite{ponnapalli2019scalable} proposed a distributed Merkle Patricia trie across multiple nodes. Extensive research has optimized Merkle tree structures~\cite{haider2018compact, ostersjo2016sparse}.
Despite these advancements, our experiments suggest that the performance benefits are limited, as trie lookup is significantly impacted by database latency.

Cache-Tries, introduced by Prokopec~\cite{hashtrie}, present a design for concurrent hash tries that ensures constant-time operations using a quiescently consistent cache.
This approach extends the cache-trie with single-linked lists, each holding pointers to nodes at one level of the trie, sorted from the deeper levels to the root.
This cache enables quick lookup and insertion by inspecting the tree from the bottom up, starting from the deepest level, or reverting to a slower top-down search from the root. 

Efficient privacy-preserving storage, as proposed by Jia et al.~\cite{blockstorage}, introduces a blockchain storage method tailored for IoT environments.
This method leverages federated extreme learning machines to classify and store hot blocks efficiently while preserving user privacy.
This approach enhances data query speed and ensures that sensitive information remains protected from unauthorized access.

Feng et al.~\cite{Feng23} propose a space-efficient storage structure for blockchain transactions that supports secure verification.
Their approach minimizes the storage footprint of transaction data while ensuring reliable verification processes.
By employing advanced data structures and algorithms, their method significantly reduces the required storage space without compromising blockchain security and integrity. 

Li et al.~\cite{lvmt} propose a replacement to the MPT data structure that, like our approach, avoids the use of an off-the-shelf key/value store, resulting in ${\mathcal O}(\log_2 n)$ lookup complexity. They instead propose a solution that utilizes a Layered Versioned Multipoint Trie data structure, resulting in ${\mathcal O}(\log n)$ lookup complexity. 

Hang et al.~\cite{hang2024slimarchive} introduced a work resembling ours. They claim that the witness proof is often not necessary in many use-cases, and propose an archive that stores only account and storage data without the MPT hashes. This easily produces a compact archive that consumes only a fraction of the full archive disk space. 

Zhang et al. \cite{zhang2024cle} proposes a new indexing mechanism for MPT based on learned model, which is smaller in size and faster for querying.

A substantial body of work has explored concurrency to address performance limitations of smart contracts~\cite{dickerson2017adding, anjana2019efficient, baheti2022dipetrans, saraph2019empirical}.
Miners construct an execution graph with forks and joins, which is replicated by other network nodes. 
This strategy removes database access from the critical path, thereby enhancing overall performance; however, it does not mitigate the latency associated with retrieving individual state values from the database.

Research has also focused on developing faster key-value databases. Faster~\cite{chandramouli2018faster} is a concurrent database that spans both hard drives and main memory with a log. Song et al.~\cite{song2018multipath} introduced bulk \verb!get! methods atop RocksDB, optimizing SSD usage. 
Papagiannis et al.~\cite{papagiannis2018efficient} reported a 4–6 times speedup over RocksDB with their memory-mapped key-value store. 
Ouaknine et al.~\cite{ouaknine2017optimization} showed substantial performance improvements with a highly-tuned RocksDB configuration on SSDs. 
Wu et al.~\cite{wu2015lsm} proposed an LSM-trie algorithm, achieving a tenfold performance enhancement with LevelDB and RocksDB, with no performance degradation as data size increased.
Lepers et al.~\cite{lepers2019kvell} introduced a database optimized for modern SSDs, while Agarwal et al.~\cite{agarwal2015succinct} suggested querying compressed data to save disk space without performance loss.

The Ethereum community is exploring sharding to distribute computational load and reduce the size of the state trie~\cite{yu2020survey}.
Sharding segments nodes into clusters by account, allowing shards to operate independently.
Recently, Ethereum has also focused on a slower L1 chain, offloading scalability to L2 chains.
\texttt{Serial index}, documented by the Klaytn Foundation~\footnote{https://docs.klaytn.foundation/docs/build/tools/indexers/}, involves the selective removal of obsolete data to optimize storage and performance. 
The Path-Based scheme, discussed in Ethereum research forums\footnote{https://ethresear.ch/t/ethereum-path-based-storage-model-and-newly-inline-state-prune/14932} and implemented in the Geth client\footnote{https://github.com/ethereum/go-ethereum/issues/25390}, introduces a mutable MPT scheme for live databases, enhancing the trie with embedded pruning capability.

Some blockchains, such as Ethereum, propose an alternative approach to handling the temporal history of blocks. The Snapshot\footnote{https://github.com/ethereum/go-ethereum/pull/20152} mechanism maintains a history of 128 blocks as a layer of flattened key-value pairs, in addition to the MPT data structure.
While this reduces the need to keep recent history in MPT, it complicates the client's software architecture, making it more error-prone (i.e., snapshots and MPT must be kept consistent), and requires extra disk space to store Snapshots.

One way to ensure that we keep the storage overheads and access time low for validators and observers is to perform live pruning~\cite{livepruning} in the MPT data structure of the \StateDB, as there are no forks and the evolution of blocks is linear. 
After a new version is introduced, the unused nodes of the old version are removed from the MPT so that only the latest worldstate of the blockchain is kept in the MPT.
Live pruning is beneficial for validators and observers, as they only require the most recent state in the \StateDB for a forkless chain. 
However, live pruning is a form of garbage collection that involves a node traversal within the MPT, collecting outdated nodes and eliminating them. 
In case of a multi-threaded client, locking must be implemented to avoid data races when live pruning is performed in parallel with new value updates, which complicates the implementation of the MPT and worsens performance.
On the other hand, \ArchiveDB does not require live pruning for forkless chains\footnote{Only for forked chains, some forks could be removed to safe storage, applying the longest-chain rule.}.
The worldstates of all blocks must be retained in the MPT. However, for forkless chains, the evolution of the versions is linear.

\section{Conclusion}
\label{sec:conclusion}
In this paper, we have presented a state database specialized for forkless blockchains. Compared to legacy state databases our solution provides improvements both in terms of throughput as well as storage overhead.  Our forkless blockchain database consists of two specialized databases, each tailored to the role of a node in a blockchain.
The \LiveDB stores data densely and allows overwriting to enable intrinsic pruning, which limits storage overhead.
The \ArchiveDB is limited to a single subsequent version per block and limits the amount of key copy overhead. Overall, we report a 100x improvement in storage and a 10x improvement in throughput compared to the geth-based Fantom Blockchain client.

\section*{Acknowledgment}
We would like to thank the following colleagues and engineers for helping with the implementation of this work, and for their valuable feedback: Jiri Malek, Jan Kalina, Lukas Pelanek.

\bibliographystyle{ACM-Reference-Format}
\bibliography{main}


\end{document}